\documentclass[%
reprint,
 amsmath,amssymb,
 aps,
superscriptaddress]{revtex4-2}

\usepackage{graphicx}%
\usepackage{dcolumn}%
\usepackage{bm}%
\usepackage{siunitx}

\begin{document}

\title{Cell mechanics, environmental geometry, and cell polarity control cell-cell collision outcomes}

\author{Yongtian Luo}
 \affiliation{William H. Miller III Department of Physics and Astronomy, Johns Hopkins University, Baltimore, Maryland 21218, USA}%
 \author{Amrinder S. Nain}
 \affiliation{Department of Mechanical Engineering, Virginia Tech, Blacksburg, VA 24061}
\author{Brian A. Camley}%
\affiliation{William H. Miller III Department of Physics and Astronomy, Johns Hopkins University, Baltimore, Maryland 21218, USA}%
\affiliation{Department of Biophysics, Johns Hopkins University, Baltimore, Maryland 21218, USA}%

\newcommand{\dif}{\mathrm{d}}
\newcommand{\vek}[1]{\boldsymbol{#1}} 

\newcommand{\add}[1]{{\color{red}{#1}}}

\begin{abstract}
Interactions between crawling cells, which are essential for many biological processes, can be quantified by measuring cell-cell collisions. Conventionally, experiments of cell-cell collisions are conducted on two-dimensional flat substrates, where colliding cells repolarize and move away upon contact with one another in ``contact inhibition of locomotion'' (CIL). Inspired by recent experiments that show cells on suspended nanofibers have qualitatively different CIL behaviors than those on flat substrates, we develop a phase field model of cell motility and two-cell collisions in fiber geometries. Our model includes cell-cell and cell-fiber adhesion, and a simple positive feedback mechanism of cell polarity. We focus on cell collisions on two parallel fibers, finding that larger cell deformability (lower membrane tension), larger positive feedback of polarization, and larger fiber spacing promote more occurrences of cells walking past one another. We can capture this behavior using a simple linear stability analysis on the cell-cell interface upon collision. 
\end{abstract}

\maketitle

\newpage

\section{\label{sec:intro} Introduction}

Collective cell behaviors play crucial roles in many physiological processes such as embryonic development and morphogenesis and pathologies such as cancer metastasis. These collective behaviors are facilitated and regulated by inter-cellular interactions \cite{Ladoux2017}. The simplest system to investigate cell-cell interactions is a pair of cells, where cell behavior after cell-cell collision can be tracked, allowing experimenters to determine the biological and physical factors regulating cell-cell interactions. One example of a stereotypical outcome of two-cell collision is contact inhibition of locomotion (CIL), during which a cell reverses its direction of motion upon contact with another cell so that both cells migrate away from each other \cite{Stramer2017}. CIL is a common behavior in cell-cell collisions for many cell types including fibroblasts and neural crest cells when they are studied on two-dimensional flat surfaces \cite{Stramer2017,Carmona-Fontaine2008}. CIL is important for cells' directional collective migration \cite{Theveneau2010,Camley2016} and pattern formation in embryogenesis \cite{Davis2012,Davis2015}. During CIL, the cell polarizes itself away from contact, e.g.\ by redistributing the activity of proteins including Rho-family GTPases such as Rac1 and RhoA \cite{Roycroft2016,Mayor2010}.  In order to increase the frequencies of cell collisions to study the rules of CIL in a more controllable way, experimentalists have used one-dimensional microchannels \cite{Lin2015} and 1D micropatterned fibronectin stripes for cells to travel along, constraining their movement and collision in straight lines \cite{Desai2013,Scarpa2013,Milano2016}. The outcomes in these collisions are dependent on various cellular properties including cell-cell adhesion, and can include reversing away from a head-on collision (e.g.\ classical CIL), cells managing to squeeze past one another to continue in their original direction (``walk-past''), or sticking together and/or forming a train of pairs of cells (``training'') \cite{Desai2013,Scarpa2013,Milano2016}. 
Cell-cell collisions also depend strongly on the geometry of the cell's environment -- as do many other cell behaviors \cite{Charras2014}. Cells on suspended nanofibers \cite{Sheets2013}, which mimic single fibers from a biologically realistic extracellular matrix \cite{Meehan2014} have drastically different collision behaviors than cells on featureless 2D substrates \cite{Singh2021}. On nanofibers, collision outcomes between cells crawling along a single fiber are predominantly walk-past, while collisions between cells attached to two parallel fibers predominantly result in training (or walk-past if they have recently divided) \cite{Singh2021}. Large rates of cells walking past other cells have also been observed for cancerous cells on micropatterns \cite{Milano2016,Bruckner2021}.

What determines collision outcomes between cells in various environments? What cellular properties and environmental factors control cell motion and interactions, giving rise to a certain cell-cell collision result? To explore these questions, we model cell behaviors during collisions in suspended fiber geometries using numerical simulations, studying the effects of mechanical properties of cells and the fiber geometry on collision outcomes. In particular, our focus is on the ``walk-past'' behavior where cells manage to push past one another, and capturing the role of the large cell shape changes that are required in this process -- processes not captured by our original self-propelled particle model \cite{Singh2021}. To model a deforming cell with a complex shape, we use a phase field approach to simulate cell motion and two-cell collisions. Phase field models have been broadly used to study cell migration at the single-cell \cite{Shao2010,Shao2012,PerezIpina2024,Zadeh2022} or collective level \cite{Nonomura2012,Camley2014,Wang2025,Loewe2020,Zhang2020,Mueller2019,Zhang2023}, including studying cell collisions on flat or micropatterned substrates \cite{Zadeh2022,Kulawiak2016}. Our model is most closely related to those of \cite{Zhang2020,Zadeh2022}. We describe the cell's polarity -- the biochemical difference between its front and back -- with a simple field driven by positive feedback, akin to \cite{Thuroff2019,Niculescu2015}. This simple model is distinct from previous modeling of cell-cell collisions using more detailed Rho GTPase polarity models \cite{Merchant2018,Kulawiak2016}. We also include cell-cell and cell-fiber adhesion (primarily facilitated by E-cadherin junctions and focal adhesions, respectively \cite{Ladoux2017}). We find we can control whether cells walk past each other or form trains with two cells moving in the same direction by tuning cell deformability, the strength of positive feedback driving cell polarity, and the distance between fibers. We capture these behaviors qualitatively with a simple linear stability analysis focused on whether cells can break symmetry at the collision to walk past one another. Our simulation results and linear stability theory, though based on a minimal modeling approach, help to reveal factors playing key roles in controlling cell-cell collisions.

\begin{figure*}[htbp]%
\centering
\includegraphics[width=\textwidth]{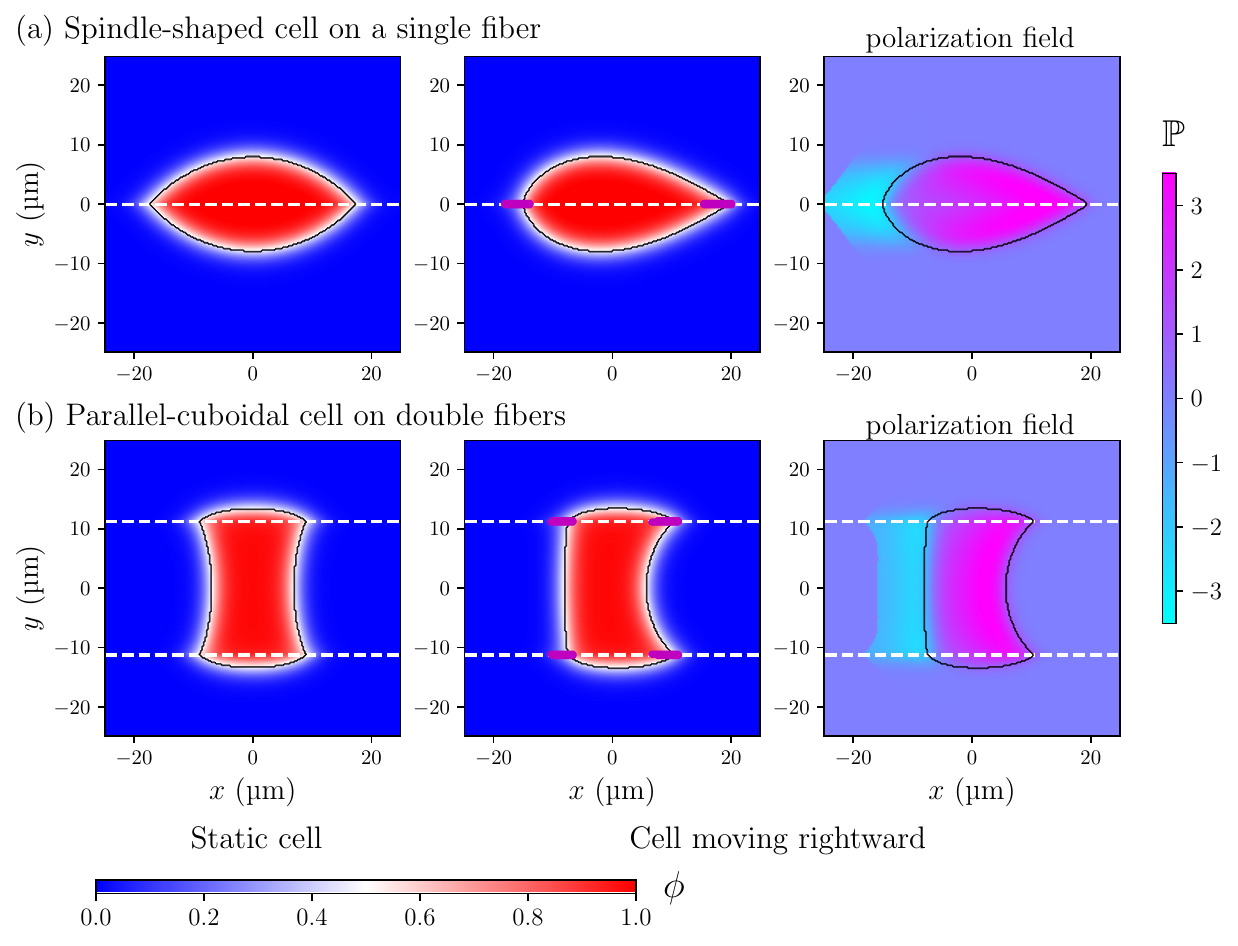}
\caption{Examples of cell phase field ($\phi$, left two columns) and polarization field ($\mathbb{P}$, the rightmost column) on suspended fibers (white dashed lines). Cells expand along fibers and become either spindle-like (a, on a single fiber) or parallel-cuboidal (b, on two parallel fibers). The cell boundary ($\phi=0.5$) is marked by a thin black contour. Left column shows cells with no active forces. Middle column shows the steady-state shapes of single cells with the full dynamic simulation. In the middle column, thick magenta line segments indicate the locations where motility forces are largest, where the absolute value $\lvert\mathbb{P}\phi^2(1-\phi)^2\chi\rvert>0.025$. The right column shows the polarity fields $\mathbb{P}$ associated with the cells in the middle column.} %
\label{single}
\end{figure*}

\section{\label{sec:model} Theoretical model and simulation method}

We use a two-dimensional phase field model to simulate pairs of cells moving along a single fiber or two parallel fibers (Fig.\ \ref{single}). 
A cell's shape is represented by a dimensionless scalar field $\phi(\vek{r}, t)$ where $\phi$ transitions continuously from one inside the cell and zero outside the cell, implicitly defining the cell boundary by $\phi=0.5$ (black contours in Fig.\ \ref{single}). We will generally work in simulation units, measuring lengths relative to the numerical lattice on which we solve our partial differential equations (with lattice size $l_0$), along with a time scale $\tau_0$, and an energy scale $\epsilon_0$. Roughly matching cell sizes and velocities as well as fiber diameters, we estimate that these scales correspond to a physical scale of $l_0$ of \SI{0.25}{\micro\m} and $\tau_0$ of \SI{0.5}{\min}. %
In this section, we first introduce free energies of cells in a fiber geometry, defining cell-substrate and cell-cell interactions, and then specify the mechanisms of cell motility and polarity that regulate cell dynamics, before briefly describing the simulation method of cell collisions.
Default values of simulation parameters can be found in Table \ref{param} and details of the numerical method can be found in Appendix \ref{app:simul}.

\subsection{Free energy functionals}

We first define cell free energy functionals that control cell shapes in our 2D phase field. We model a cell as having a characteristic tension opposing extension of its membrane (a line tension in our 2D model), a constraint on area, and an adhesive energy promoting its spreading along the nanofibers. As a result, the free energy functional for a phase field $\phi_i(\vek{r}, t)$ of a single cell $i$ is:
\begin{widetext}
\begin{equation}
\label{fi}
F_i=\int\dif\vek{r}\frac{\alpha}{4}\Big(\phi_i^2(1-\phi_i)^2+d^2(\nabla\phi_i)^2\Big)+\lambda\Big(1-\int\dif\vek{r}\frac{\phi_i^2}{\pi R_0^2}\Big)^2-\frac{E_0 A_{\text{ad}}(\phi_i)}{A_{\text{ref}}+A_{\text{ad}}(\phi_i)}.
\end{equation}
\end{widetext}
The first term in Eq.\ \eqref{fi} is the Cahn-Hilliard free energy \cite{Loewe2020,Zhang2020,Zadeh2022}. The parameter $\alpha$, which is related to cell membrane tension, has units $\epsilon_0/l_0^2$. The Cahn-Hilliard term enforces the phase field to have $\phi_i=1$ inside cell $i$ and $\phi_i=0$ outside the cell, and penalizes the formation of a longer cell boundary. The characteristic length $d$ sets the cell interface width. %
  The line tension of the cell boundary is $\Gamma$ (units of energy/length, $\epsilon_0/l_0$), given by \cite{Loewe2020}%
\begin{equation}
\label{lin}
\Gamma=\frac{2}{3}d\alpha=\frac{4\sqrt{5}}{3}l_0\cdot\alpha.
\end{equation}
The line tension $\Gamma$ is directly proportional to $\alpha$, so in this work, we vary $\alpha$ to control line tension. The second term of Eq.\ \eqref{fi} is a soft constraint on the cell area $\int\dif\vek{r}\phi_i^2$, which penalizes deviations from the area of a circle with radius $R_0$. The cell radius is $R_0=40l_0=\SI{10}{\micro\m}$, roughly reasonable for many eukaryotic cells. If only line tension and cell area constraint were taken into account, a free, static cell would evolve to a circle of radius $R_0$. %

The third term in Eq.\ \eqref{fi} models the adhesive energy between the cell body and the fiber substrate with an adhesion strength $E_0$ \cite{Bischofs2008,Albert2014}. This negative energy promotes the expansion of the area of adhesive contact between cell $i$ and the fiber $A_{\text{ad}}(\phi_i)$.  This adhesion energy saturates to its minimum value $-E_0$ as the adhesive area exceeds the scale $A_{\text{ref}}=1l_0^2$ reflecting the finite concentration of adhesive molecules (e.g., proteins like integrin that mediates the formation of focal adhesions \cite{Ladoux2017,Dubin-Thaler2004}). We define the cell-fiber adhesive area in Eq.\ \eqref{fi} as:
\begin{equation}
\label{ad}
A_{\text{ad}}(\phi_i)=\int\dif\vek{r}\phi_i^2(3-2\phi_i)\chi
\end{equation}
where we have introduced a separate phase field $\chi(\vek{r})$ that indicates the locations of fibers, which are present wherever $\chi=1$ and absent elsewhere ($\chi=0$). $\chi$ is nonzero over a width of \SI{500}{\nano\m} (two lattice spacings $l_0$), which matches the diameter of nanofibers used in the experiments of \cite{Singh2021}. The expression in Eq.\ \eqref{ad} measures the overlapping area of cell-fiber contact where both $\phi_i$ and $\chi$ are nonzero. %
We use the somewhat nonobvious definition of Eq.\ \eqref{ad} to avoid numerical issues.
This form of $A_{\text{ad}}$ has a functional derivative $\delta A_{\text{ad}}/\delta\phi_i=6\phi_i(1-\phi_i)\chi$ -- we can see that changes in the cell boundary, where $\phi_i(1-\phi_i)$ is nonzero, change the adhesive area, as we would expect. %
 Our initial idea was using a simpler form $A_{\text{ad}}^\textrm{simple}\equiv\int\dif\vek{r}\phi_i\chi$; we found this choice to lead to issues where $\phi_i$ has a low nonzero value along the fibers even far from the cell boundary. 

The combined effect of our line tension and cell-fiber adhesion is that single cells that are initially circular and have no active driving forces (left panel of Fig.\ \ref{single}) spread along the fibers (indicated by dashed white lines), but come to a steady state determined by the competition of line tension, which tends to promote a circular shape, and cell-substrate adhesion, which tends to elongate the cell along the fibers. 

Cell-cell interactions are described by an additional free energy functional for cells $i$ and $j$ \cite{Camley2014,Zadeh2022}:
\begin{equation}
\label{fij}
F_{i,j}=\int\dif\vek{r}\Big(\frac{g}{2}\phi_i^2\phi_j^2-\frac{\sigma}{4}\phi_i^2(1-\phi_i)^2\phi_j^2(1-\phi_j)^2\Big)
\end{equation}
which includes the effects of both cell-cell repulsion (the exclusion of cell area overlap where both $\phi_i$ and $\phi_j$ are one, with an excluding strength $g$), and the adhesion with strength $\sigma$ at cell-cell interface where cell boundaries ($\phi(1-\phi) \neq 0$) touch. %
The total free energy functional is:
\begin{equation}
\label{ftot}
F=\sum_i\Big(F_i+\sum_{j\neq i}F_{i,j}\Big)
\end{equation}
where $i,j=1$ or 2.

\subsection{Dynamics of the phase field model}

Here we introduce the motility mechanisms that drive cell motion. The phase field $\phi_i$ evolves by \cite{Zhang2020}:
\begin{equation}
\label{evo}
\frac{\partial\phi_i}{\partial t}=-\Big(\vek{v_i}\cdot\nabla\phi_i+\frac{1}{\gamma}\frac{\delta F}{\delta\phi_i}\Big)
\end{equation}
which includes a velocity field $\vek{v_i}(\vek{r}, t)$ arising from the cell's self-propulsion, which advects the cell boundary, as well as the second term on the right hand side, which describes the relaxation of cell shape to the local minimum of $F$. %

The velocity field $\vek{v}_i$ for each cell is created by an active motility force density $\vek{f_i}(\vek{r}, t)$ (force per unit area, with units $\epsilon_0/l_0^3$) arising from actin polymerization and myosin contraction on cell $i$. We assume that the relationship between these arises from the overdamped limit of viscous fluid flow \cite{Zhang2020}, leading to:
\begin{equation}
\label{vi}
\vek{v_i}=\frac{1}{\eta}\Big(\frac{\delta F}{\delta\phi_i}\nabla\phi_i+\vek{f_i}\Big)+\vek{\tilde{v}_i}.
\end{equation}
The first term on the right hand side comes from a force balance between the friction force density $-\eta\vek{v_i}$ (with a friction coefficient $\eta$) and the combination of the passive force density $(\delta F/\delta\phi_i)\nabla\phi_i$ (due to cell shape relaxation) and active motility force density $\vek{f_i}$. Because cell-cell collisions are stochastic, we have added a velocity noise term $\vek{\tilde{v}_i}$, which is sampled and implemented every $t_\textrm{sample}$ onto the cell velocity field. This term is uniform in space, modeling fluctuations in the whole cell's velocity. We assume the added noise is Gaussian, {$\vek{\tilde{v}_i}=\tilde{v}_0(X\vek{\hat{x}}+Y\vek{\hat{y}})$} drawing independent random variables $X\sim\mathcal{N}(0,1)$ and $Y\sim\mathcal{N}(0,1)$ at each sampling time point $t_\textrm{sample}$. This random velocity is held constant for a period of $t_\textrm{sample}$ until the next sampling time point. We choose $t_\textrm{sample} = 0.002\tau_0$ which happens to be equal to the simulation time step $\Delta t$ used in our algorithm, but note that it should not vary if $\Delta t$ is changed; see Appendix \ref{app:simul}. With our typical range of parameters (those in Table \ref{param} and used in the plots of Fig.\ \ref{L90}--\ref{L90s0}), we obtain average cell centroid speeds around 0.2 to $\SI{2}{\micro\m}/\SI{}{\min}$, depending on the specific parameters chosen. This range of cell speeds is roughly consistent with that seen in experiments \cite{Singh2021}. We note the mean cell speeds are much larger than the velocity noise strength $\tilde{v}_0=0.1l_0/\tau_0=\SI{0.05}{\micro\m}/\SI{}{\min}$ -- our noise here is a small perturbation, meant to break symmetry rather than realistically represent the degree of variability in cell speeds observed in \cite{Singh2021}. 

\subsection{Cell polarization field}

Where the cell exerts protrusive vs.\ contractile force will depend on the cell's polarity -- what part of the cell is the ``front.'' We characterize this polarity using the dimensionless scalar field $\mathbb{P}_i(\vek{r}, t)$ of cell $i$. This single field $\mathbb{P}_i$ is a summary of multiple biochemical processes and molecular details that characterize a cell's polarity, including the distributions of Rho GTPases like Rac1 and RhoA, as well as actin and myosin, centrosome orientation, etc.\ \cite{Ladoux2017}. 
Our convention is that protrusion occurs where $\mathbb{P}_i>0$ (cell front) and contraction occurs where $\mathbb{P}_i<0$ (cell rear). We assume that cell-substrate contact is necessary for force generation \cite{chan2008traction,xia2008directional}, so force density $\vek{f_i}$ is localized at the overlaps between cell boundary and fiber substrates (also the typical locations of focal adhesions \cite{Jana2019}). To localize force to this region, we need both $\chi=1$ and $\phi_i^2(1-\phi_i)^2$ to be nonzero, suggesting:
\begin{equation}
\label{ffi}
\vek{f_i}=A\mathbb{P}_i\phi_i^2(1-\phi_i)^2\chi\vek{n_i}
\end{equation}
where $A$ sets the strength of protrusion and $\vek{n_i}=-\nabla\phi_i/\lvert\nabla\phi_i\rvert$ is the outward normal unit vector to cell $i$.

We want to capture persistent cell migration. We use a very simple positive feedback mechanism, akin to those proposed by \cite{Niculescu2015,Thuroff2019} in which successful protrusions promote further protrusion (increasing polarity). Within our phase field framework, we thus promote increases in polarity $\mathbb{P}_i$ where the cell has already started to protrude -- the regions where $\partial\phi_i/\partial t>0$, and decrease $\mathbb{P}_i$ where contractions occur $\partial\phi_i/\partial t<0$. The simplest implementation of this idea is:
\begin{equation}
\label{ppi}
\frac{\partial\mathbb{P}_i}{\partial t}=\beta\phi_i\frac{\partial\phi_i}{\partial t}-\frac{\mathbb{P}_i}{\tau}
\end{equation}
where $\beta$ is a dimensionless parameter that sets the positive feedback strength, and $\tau$ sets the timescale over which polarity decays. We adopt, for simplicity, the convention that polarity evolves in the reference frame of the substrate -- appropriate if polarity is tightly associated with the basal membrane or with focal adhesions -- so that there is no velocity advecting $\mathbb{P}_i$. While the polarization field $\mathbb{P}_i$ is created only interior to the cell, where $\phi_i > 0$, it may persist transiently after the cell leaves, decaying with time constant $\tau$; see the right column of Fig.\ \ref{single}. In these images, for the sake of transparency, we show the polarity field in the whole area, while only the polarity near the cell boundary actually leads to any effect in the model. When evolving the polarity dynamics of Eq.\ \eqref{ppi}, we set $\mathbb{P}_i = 0$ far from the cell (where $\phi < 0.001$); see Appendix \ref{app:simul}. 

\begin{table}[hbt!]
\centering
\begin{tabular}{|c|c|c|}
\hline
Parameter & Value & Location \\ \hline
Length unit (lattice size) $l_0$ & \SI{0.25}{\micro\m} & \\
Time unit $\tau_0$ & \SI{0.5}{\min} & \\ \hline
Cell boundary width $d$ & $\sqrt{20}l_0$ & Eq.\ \eqref{fi}\\
Cell initial radius $R_0$ & $40l_0$ & \\
Cell area constraint strength $\lambda$ & $1000\epsilon_0$ & \\
Cell-fiber adhesion strength $E_0$ & $2000\epsilon_0$ & \\
Cell adhesion reference area $A_{\text{ref}}$ & $1l_0^2$ & \\ \hline
Cell-cell area exclusion strength $g$ & $1\epsilon_0/l_0^2$ & Eq.\ \eqref{fij}\\
Cell-cell adhesion strength $\sigma$ & $5\epsilon_0/l_0^2$ & \\ \hline
Transport coefficient $1/\gamma$ & $50l_0^2/(\epsilon_0\tau_0)$ & Eq.\ \eqref{evo}\\ \hline
Friction coefficient $\eta$ & $1\epsilon_0\tau_0/l_0^4$ & Eq.\ \eqref{vi}\\
Cell velocity noise strength $\tilde{v}_0$ & $0.1l_0/\tau_0$ & \\ \hline
Motility force strength $A$ & $200\epsilon_0/l_0^3$ & Eq.\ \eqref{ffi} \\ \hline
Polarization decay time constant $\tau$ & $100\tau_0$ & Eq.\ \eqref{ppi}\\ \hline
Simulation time step $\Delta t$ & $0.002\tau_0$ & Eq.\ \eqref{phi}\\
\hline
\end{tabular}
\caption{\label{param} Summary of default parameter values used in simulations.}
\end{table}

\subsection{Simulation method}

We solve the phase field and polarity equations Eq.\ \eqref{evo} and Eq.\ \eqref{ppi} using a finite difference method (Appendix \ref{app:simul}).  To start a simulation trajectory, we initialize the phase field system by setting a circular cell of radius $R_0$ on the fibers and allowing the cell to spread over the fibers without active forces, obtaining a static ``relaxed'' cell shape. These are shown in the left panel of Fig.\ \ref{single}. We then simulate the free motion of a single isolated cell by driving the obtained static cell to move along the fibers while allowing the polarity field to evolve normally by Eq.\ \eqref{ppi} starting with an asymmetric polarization field. The cell in motion reaches a steady state with a stable shape after long enough time. We use such dynamic steady states as initial cell conditions when simulating two-cell collisions, ensuring that the cells involved in a collision are generated from identical parameters and already stabilized on fibers (having traveled for sufficiently long time and not affected by any particular initial state we choose for the simulation) prior to the collision. We simulate two-cell collisions under various conditions, whose outcomes depend on cellular properties and characteristics of the fiber environment. More simulation details can be found in Appendix \ref{app:simul}.

\begin{figure*}[hbt!]
\centering
\includegraphics[width=\textwidth]{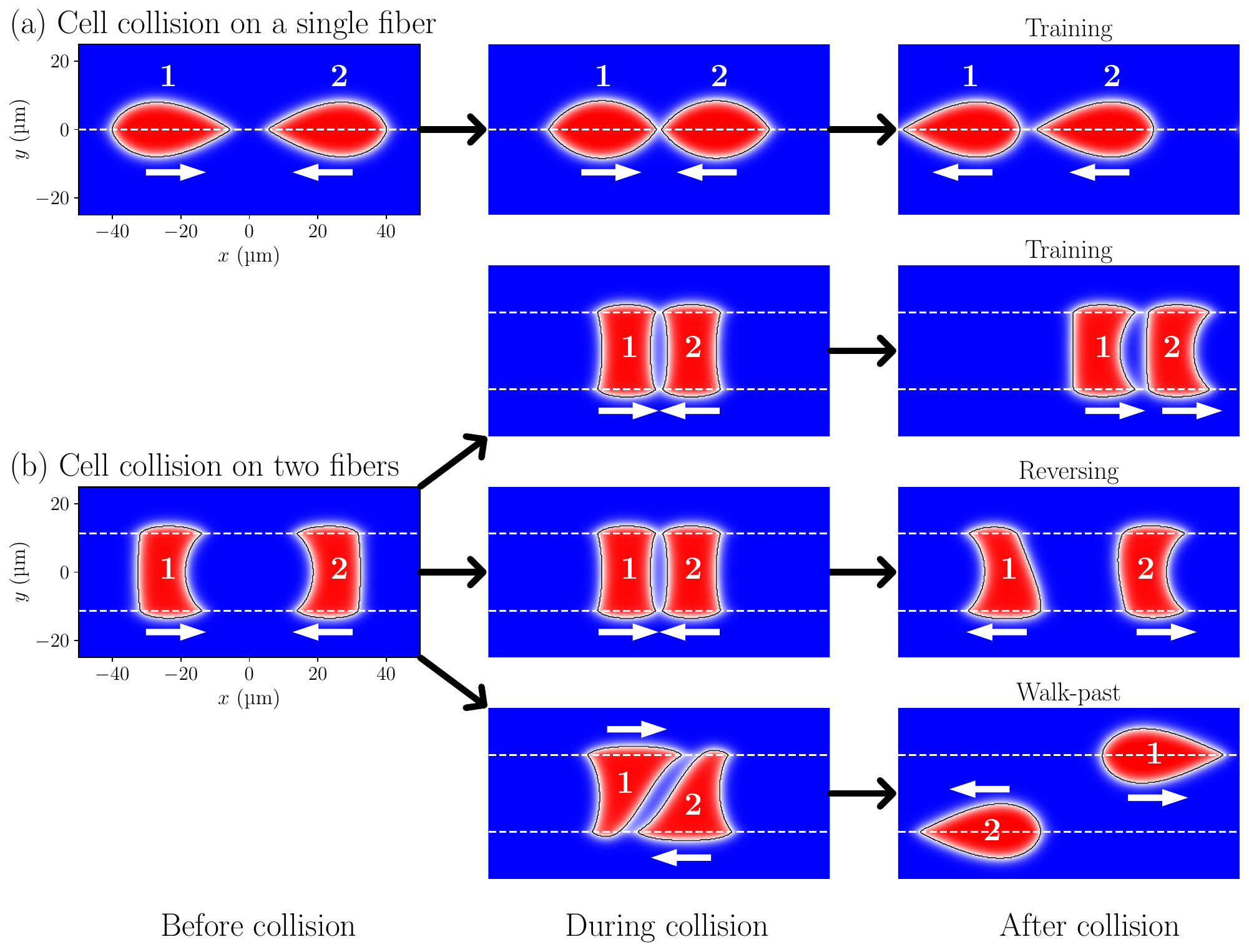}
\caption{Possible collision outcomes of cells on fibers. Directions of motion are marked by white arrows. (a) In our simulations, the collision of two spindle-shaped cells on a single fiber almost always results in cells sticking together and moving in the same direction (\lq\lq training\rq\rq). (b) The collision of two parallel-cuboidal cells on two fibers can also result in training, or possibly reversing from each other akin to classical CIL, which is most likely to occur in simulations without cell-cell adhesion. In addition, cells can walk past each other while keeping their original directions of motion. During a walk-past, both cells shrink to a single fiber and become teardrop-like. In other collision outcomes (training and reversing), one or both of the cells can also shrink to single-fiber shape (not shown in this figure). {Cell-cell collision movies for common outcomes and a few unusual cases can be seen in Movies 1-8.}}
\label{double}
\end{figure*}

\section{\label{sec:result} Results}

If we simulate single cells on a single fiber, we obtain spindle-shaped cells (Fig.\ \ref{single} (a)), consistent with the experiments of \cite{Singh2021}. Cells simulated on two parallel fibers take on a near-rectangular shape (e.g.\ in Fig.\ \ref{single} (b) where fibers are separated by spacing $L=90l_0$ or \SI{22.5}{\micro\m}, which is comparable to cell sizes in this work and fiber spacings in \cite{Singh2021}) -- we call these cells ``parallel'' or ``parallel-cuboidal'' as in the experimental work of \cite{Singh2021}.  %

What happens if we collide two cells together? We find that if we collide cells on a single fiber (spindle cells), they almost always form a train (nearly 100\% probability at our default parameters) with one cell moving in its original direction of motion and forcing back the other cell which repolarizes and turns around (Fig.\ \ref{double} (a), Movie 1). Experimentally, we earlier found that spindle cells most commonly walk past one another \cite{Singh2021} -- but we would not expect to see this in our two-dimensional model, because for cells attached to single fibers, the walk-past always involves cells moving past each other in the third dimension. Instead, we see training, which was the second-most-common experimental outcome for spindle collisions in \cite{Singh2021}. Because of this expected discrepancy, we do not focus on spindle-spindle collisions in this paper. Parallel-cuboidal cells are more planar in 3D real space and are better captured by the 2D phase field, so they are our primary interest here.

If we simulate collisions of parallel cells attached to two fibers (also separated by $L=\SI{22.5}{\micro\m}$ in Fig.\ \ref{double} (b)), we find, depending on the parameters, multiple distinct outcomes, including training, walk-past, and rarely, cell reversals. This is consistent with the  behavior of fibroblasts on parallel nanofibers, which we earlier found to either form trains with near-100\% probability (cells that have not recently divided) or most commonly walk past one another (cells that have recently divided, which are faster, and cells in low calcium media) \cite{Singh2021}. %
In a training event, cell pairs spontaneously break symmetry and end up moving together either leftward or rightward (see Movie 2). In a walk-past, in order to keep their original opposite directions of motion and avoid each other, the two cells initially attached to both fibers will continue protruding on different single fibers along their respective directions upon contact (as in the last row of Fig.\ \ref{double}). Each cell gradually shrinks to attach only onto the fiber where it continues protruding and detaches from the other fiber, and becomes teardrop-like in shape on single fiber (as spindle cells in motion, see Movie 3). The cells are in steady motion on single fibers after collision and do not resume the initial two-fiber parallel shapes. In the least likely outcome of reversing, both cells repolarize and turn around as in a classical CIL (see Movie 4). In most cases, cells in training or reversing keep their parallel shapes throughout the process, but in a few cases one cell (see Movies 5 and 6) or both cells (see Movies 7 and 8) could shrink to single fiber (similar to cells experiencing walk-past). %

What factors control the outcomes of cell-cell collisions? We study the effects of three varying parameters on collision outcomes between parallel cells on two fibers: the line tension parameter $\alpha$ (which is proportional to line tension $\Gamma$, characterizing the softness or deformability of a cell), the positive feedback strength $\beta$ of cell polarity, and the spacing $L$ between parallel fibers. We sweep the parameters $(\alpha,\beta)$ over the ranges of $0.1\epsilon_0/l_0^2\leqslant\alpha\leqslant 0.2\epsilon_0/l_0^2$ with an interval of $\Delta\alpha=0.02\epsilon_0/l_0^2$ and $10\leqslant\beta\leqslant 50$ with an interval of $\Delta\beta=4$, and simulate three fiber spacings $L=90l_0$, $80l_0$ and $70l_0$ (\SI{22.5}{\micro\m}, \SI{20}{\micro\m} and \SI{17.5}{\micro\m}, respectively, similar to experimental values from \cite{Singh2021}). For each of these parameter sets, we perform 96 independent simulation trajectories and obtain the statistics of the number of each possible collision consequence, identifying the dominant outcome and analyzing both qualitative and quantitative effects of the parameters. These results are shown in Fig.\ \ref{L90}--\ref{L70}. %
For each of these figures, the phase diagram in (a) illustrates the dominant collision outcomes. We call an outcome dominant if it occurs at least 53 times out of 96 simulations (or 55\% of the time), while we also show the coexistence of significant presence of both walk-past and training (labeled by superposition of symbols in phase diagrams) when the difference of frequencies between the two outcomes is less than 10 (roughly 10\%).

\subsection{Effect of line tension, positive feedback strength, and fiber spacing on cell-cell collision outcomes}

\begin{figure}%
\centering
\includegraphics[width=\columnwidth]{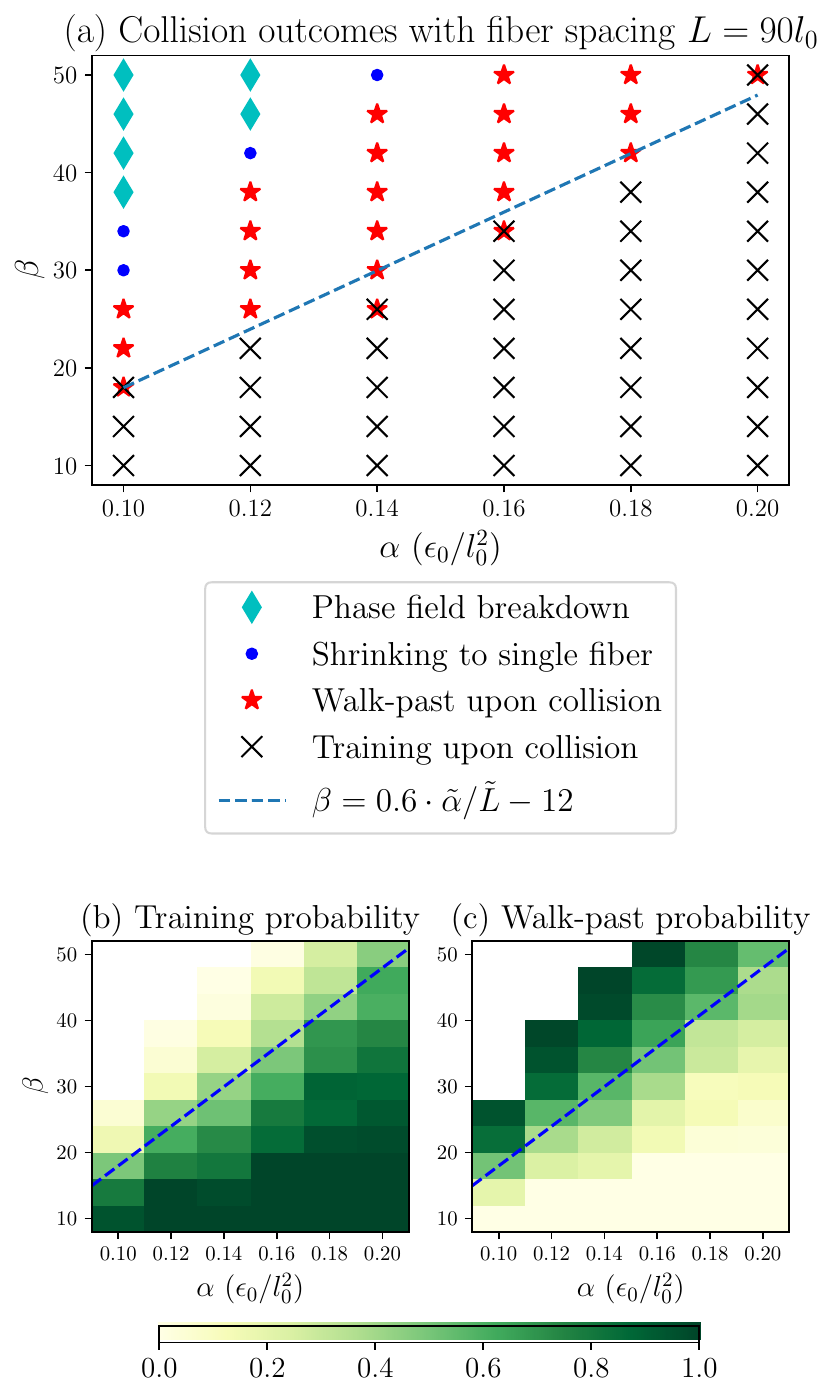}
\caption{(a) Phase diagram showing dominant cell collision outcomes on two parallel fibers separated by a spacing $L=90l_0$ {(\SI{22.5}{\micro\m})}. %
 (b) and (c) show the frequencies of training and walk-past results, respectively, calculated from 96 independent simulation trajectories at each point in the phase diagram. The blue dashed line in all three figures is a global fit to the linear stability analysis predicting the transition between training and walk-past.}
\label{L90}
\end{figure}

We show the results of our simulations on the widest fiber spacing $L = 90 l_0$ in Fig.\ \ref{L90}. Most outcomes of these collisions are either walk-past events (stars) or training events (crosses) in Fig.\ \ref{L90} (a). There is a transition between primarily training outcomes (at high $\alpha$ -- i.e.\ high line tension -- and low positive feedback strength $\beta$) and primarily walk-past (high $\beta$ and low $\alpha$). More quantitatively, the probability of training and walk-past at each parameter set is shown in Fig.\ \ref{L90} (b)--(c). This transition from training to walk-past at high $\beta$ and low $\alpha$ makes intuitive sense. The force driving the cells to walk past one another depends on the positive feedback $\beta$ -- more polarized cells are more persistent and more likely to walk past. However, to walk past, cells must deform, which is resisted by the cell's line tension $\Gamma \sim \alpha$. We will later formalize this intuition in a linear stability theory in Subsection \ref{lsa}, which predicts the shape of the dashed line in Fig.\ \ref{L90}. We also find that at sufficiently small $\alpha$ ($\leqslant 0.14 \epsilon_0/l_0^2$ in the range of the phase diagram) and large enough $\beta$, parallel cells are themselves not stable. A cell initially attached to two fibers will become unstable and unable to keep hanging onto both fibers under the strong driving force, and shrink to a single fiber even without interacting with another cell; these cases are indicated by blue dots in Fig.\ \ref{L90} (a). An example of this behavior is shown in Movie 9. With even smaller $\alpha$ or larger $\beta$, cell phase fields break apart and fail to simulate cell behaviors normally (as labeled by cyan diamonds). In these cases, cells experience large-scale deformations, such as being stretched along a fiber to a large extent while narrowing in the perpendicular direction so that the cell interior thickness is no longer larger than the finite boundary width $d$ used in our phase field model. This sort of behavior would be expected in the limit of zero line tension, when the cell would prefer to completely wet the substrate.

\begin{figure}%
\centering
\includegraphics[width=\columnwidth]{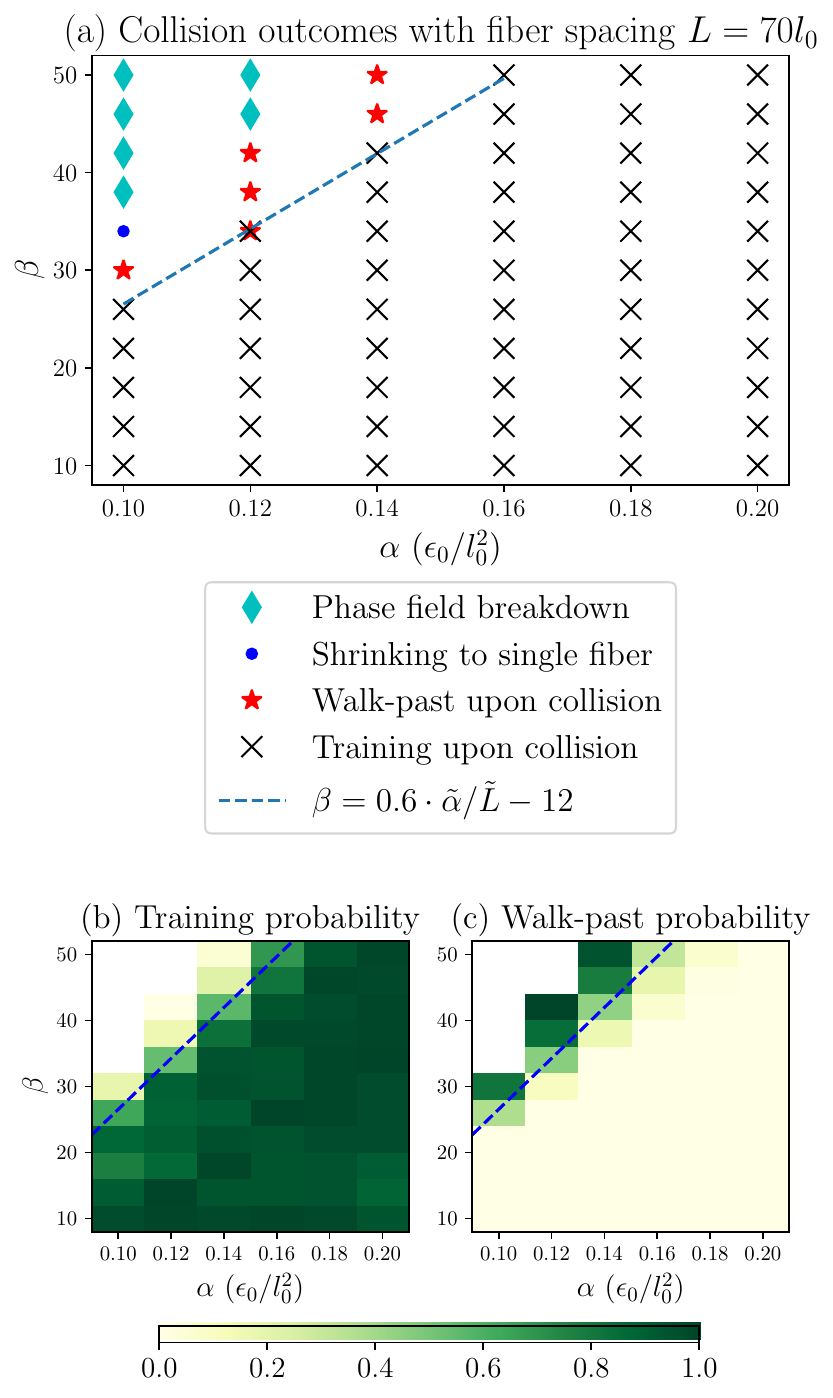}
\caption{(a) Phase diagram showing dominant cell collision outcomes on two parallel fibers separated by a spacing $L=70l_0$ {(\SI{17.5}{\micro\m})}. (b) and (c) show the frequencies of training and walk-past results calculated from 96 independent simulation trajectories at each point in the phase diagram. Note that compared to the case of $L=90l_0$, more points are dominated by training and only a few are dominated by walk-past. The linear stability prediction of walk-past threshold (dashed blue line) also moves upward accordingly.}
\label{L70}
\end{figure}

How do cell-cell collisions change when we change the fiber spacing $L$? We show outcomes of cell-cell collision simulations at narrower fiber spacing ($L = 70 l_0$) in Fig.\ \ref{L70}. Comparing with Fig.\ \ref{L90}, we see that training is far more common at smaller fiber spacing -- the transition between training and walk-past moves toward smaller $\alpha$ and larger $\beta$. This shift in the transition line is predicted by our linear stability analysis in Subsection \ref{lsa}, given by the blue dashed line in Fig.\ \ref{L70}. %
Given our earlier intuition, we think of decreasing the fiber spacing $L$ as effectively making cells harder to deform, with similar effects as increasing the line tension (or $\alpha$). Unsurprisingly, we find that collision outcomes with spacing $L = 80 l_0$ are intermediate between $L = 70 l_0$ and $L = 90 l_0$ (Fig.\ \ref{L80} in the appendix.)

\subsection{\label{noadh} Collision outcomes of two cells on parallel fibers with zero cell-cell adhesion}

\begin{figure}%
\centering
\includegraphics[width=\columnwidth]{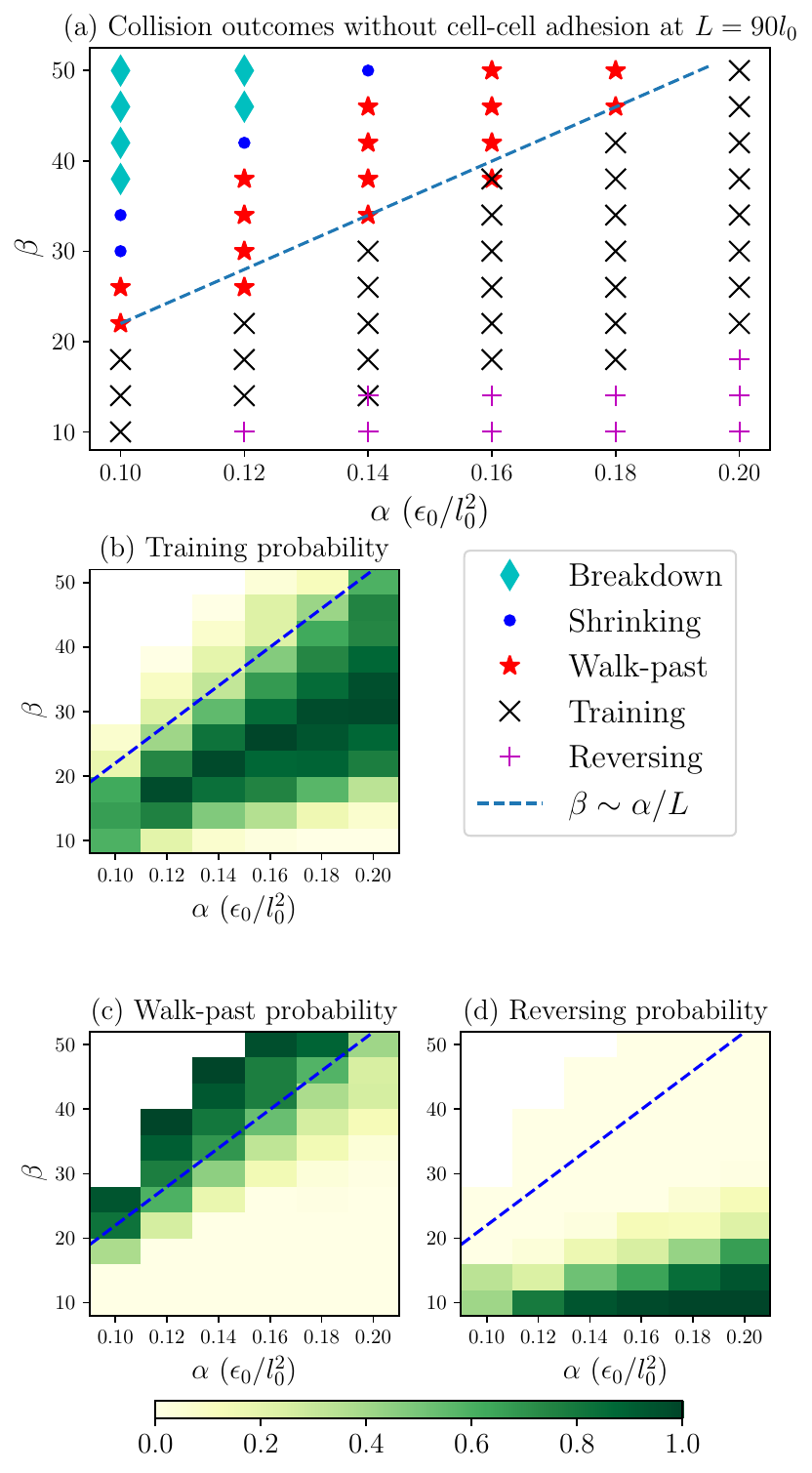}
\caption{(a) Phase diagram showing dominant cell collision outcomes without cell-cell adhesion, on two parallel fibers separated by a spacing $L=90l_0$ {(\SI{22.5}{\micro\m})}. The absence of cell boundary adhesion leads to the dominance of both cells reversing from each other upon collision at large $\alpha$ and small $\beta$. The linear stability prediction of walk-past threshold (dashed line) moves slightly to the left compared to the one at nonzero cell-cell adhesion in Fig.\ \ref{L90}. (b), (c) and (d) show the frequencies of training, walk-past, and reversing results, respectively, calculated from 96 independent simulation trajectories at each point in the phase diagram. }%
\label{L90s0}
\end{figure}

Experimentally, fibroblasts in low-calcium media (which have low cell-cell adhesion due to the calcium-dependence of cadherin interactions) were found to almost always walk past one another in parallel collisions \cite{Singh2021}. However, our physical intuition for our model suggests a different outcome -- cell-cell adhesion will {\it decrease} the energy per unit length of the cell-cell contact, making it easier for one cell to walk past another. We test this idea by simulating collisions between cells with $\sigma = 0$, removing cell-cell adhesion entirely (Fig.\ \ref{L90s0}). %
We find that outcomes are mostly similar to the results with nonzero cell-cell adhesion (compared with Fig.\ \ref{L90}), but can be qualitatively different in some regions of the phase diagram. Most dramatically, we see that cell reversal (classical CIL) emerges as the dominant outcome of collision (labeled by plus signs) with large line tension parameter $\alpha$ and small $\beta$. In this parameter range, cells are least likely to deform and training would be the dominant result with nonzero $\sigma$. This suggests that cell-cell adhesion helps keep cells stuck together until they come to a consensus direction and form a train. We also suspect that, similarly, if we disrupted cell-cell adhesion in experiments of fibroblast collisions in the spindle case, we could convert relatively common training events into reversals. 

What happens to the transition boundary between training and walk-past events? At smaller $\alpha$ and/or larger $\beta$ across the central part of the diagram, the collision outcomes are dominated by training, similar to Fig.\ \ref{L90}, but with slightly more training and fewer walk-past results in the upper left half of the diagram -- the absence of cell-cell adhesion shifts the transition line up and to the left on the graph. This small shift of the transition between walk-past and training is consistent with the relatively small effect of cell-cell adhesion on the interfacial energy, and can be incorporated into our linear stability analysis (see Subsection \ref{lsa} and Appendix \ref{app:shift}). The shift of the blue dashed line can thus be predicted again using the linear stability analysis. Subplots in Fig.\ \ref{L90s0} (b), (c) and (d) display the distributions of probabilities of training, walk-past, and reversing respectively, also calculated from 96 independent simulation results at each point. 

We found that decreasing cell-cell adhesion increases training and decreases walk-past -- in conflict with experimental observations of fibroblasts in a low-calcium environment which reduces cell-cell adhesion and leads to walk-past almost 100\% of the time after collisions \cite{Singh2021}, but consistent with our physical intuition. We discuss potential explanations for this discrepancy later in the Discussion (Section \ref{sec:disc}).%

\subsection{\label{lsa} Linear stability analysis}

\begin{figure*}[hbt!]
\centering
\includegraphics[width=\textwidth]{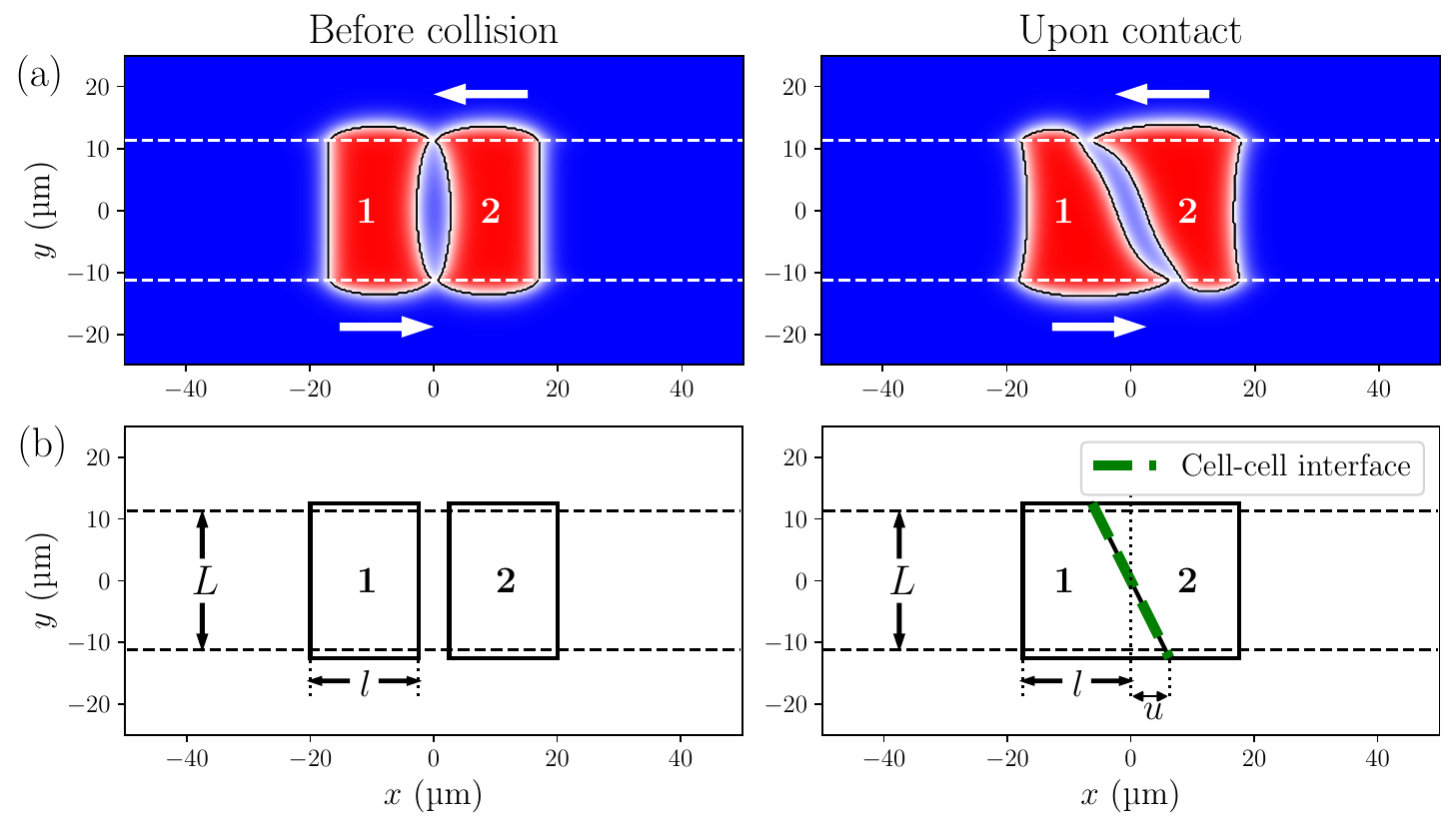}
\caption{Illustration of two cells on parallel fibers approaching each other and forming a cell-cell interface upon contact. Before the collision, the shape of both cells 1 (left) and 2 (right) represented by phase fields in (a) can be approximated by a rectangle of height $L$ (the fiber spacing) and length $l$ (along fibers) in (b). Upon collision, with the formation of interface between cell boundaries (thick green dashed line in (b)), the length of one side of a rectangular cell on a fiber extends to $l+u$ and its other side on the other fiber shortens to a length of $l-u$, contributing to the increase of perimeter of each cell.}
\label{lst}
\end{figure*}

In an attempt to quantify the locations of transition between walk-past and training in the phase diagrams in Fig.\ \ref{L90}--\ref{L90s0}, we develop a simple linear stability theory for when walk-past is possible. Our core idea is that a necessary -- though not necessarily sufficient -- condition for walk-past to occur is that  upon an initially symmetric cell-cell contact, the symmetric state should be unstable, leading to the cell-cell interface tilting (illustrated in Fig.\ \ref{lst}). The growth of this tilted interface is penalized by factors like the line tension, which should suppress the development of the instability at cell-cell contact.

We first consider the energetic cost for the deformation of a cell attached onto parallel fibers separated by spacing $L$, assuming this energy is dominated by the line tension $\Gamma$. Initially, the cell shape resembles a rectangle of height $L$ and length $l$  (the left images of Fig.\ \ref{lst}). The perimeter of this rectangular cell is $2(L+l)$.  Upon collision with another cell, we see that the boundary of the cell-cell contact tilts and elongates. We approximate this by assuming a near-linear contact shape, with the length of one horizontal boundary segment along one fiber slightly elongating to $l+u$ and the horizontal segment along the other fiber shortening to $l-u$. The length of the cell-cell contact thus extends from its initial length $L$ to $\sqrt{L^2+4u^2}$. The change of perimeter compared to an undeformed cell is $\sqrt{L^2+4u^2}-L$. At the onset of collision we suppose $u\ll L$, and when given the line tension $\Gamma$ in Eq.\ \eqref{lin}, we have the extra energetic cost $H$ of forming the stretched interface approximated as (for each cell):
\begin{equation}
\begin{split}
H&=\Gamma(\sqrt{L^2+4u^2}-L)\\
&=\Gamma L\Big[\sqrt{1+4\Big(\frac{u}{L}\Big)^2}-1\Big] \\
&\approx\Gamma L\Big[1+2\Big(\frac{u}{L}\Big)^2-1\Big] \\
&\approx\frac{2\Gamma}{L}u^2.
\end{split}
\end{equation}

Our full phase field model has an overdamped dynamics in the absence of motility. If there were no active forces, the boundary would simply relax back to its equilibrium shape. Because of this, we expect an overdamped relaxation of the deformation $u$, i.e.\ $\frac{\partial }{\partial t}u(t) = -a \frac{\partial H}{\partial u}$ with $a >0$ a constant prefactor. Competing against this relaxation is the polarity model, which is driven by positive feedback -- successful protrusion leads to more protrusion. This suggests that there should be a term that drives positive feedback in $u$, which promotes the growth of the cell-cell interface. We assume this term can be approximated by, at small $u$, some positive function $f(\beta) u$, leading to 
\begin{equation}
\frac{\partial u}{\partial t}=-a \frac{\partial H}{\partial u}+f(\beta)u=\Big(f(\beta)-4a\frac{\Gamma}{L}\Big)u.
\end{equation}
When $f(\beta)<4a\Gamma/L$, any initial deformation of the interface $u$ shrinks with time -- $u = 0$ is the stable solution, making a walk-past unlikely to happen as an outcome of the collision. When $f(\beta)>4a\Gamma/L$, the straight interface deformation $u = 0$ is linearly unstable, and we see that cells will break symmetry at the cell-cell contact. This is a necessary ingredient for one cell to go to the top fiber and one to the bottom, as occurs in a walk-past event. If the function $f(\beta)$ is linear, we would then predict that walk-past is only possible once $\beta > \beta_\textrm{crit}$, with $\beta_\textrm{crit}$ a linear function of $\Gamma/L$, or equivalently $\alpha/L$.

Our linear stability theory predicts that walk-past occurs if $\beta$ is above a critical value, and that this critical value increases with $\Gamma$ (which is proportional to $\alpha$) and decreases with fiber spacing $L$ -- this matches our physical intuition, and is qualitatively consistent with our results in Fig.\ \ref{L90} and \ref{L70}. Can we make this more quantitative? Because we have not linked $a$ or the linear function $f(\beta)$ to our simulation parameters, we must set these from our simulation results. We find that the value $\beta_\textrm{crit}$
\begin{equation}
\label{bet}
\beta_\textrm{crit}=0.6\frac{\tilde{\alpha}}{\tilde{L}}-12
\end{equation}
provides a good global description of the transition between walk-past and training in Fig.\ \ref{L90} and \ref{L70} (and also \ref{L80}) -- this is the blue dashed line. We write this equation in terms of the nondimensionalized parameters $\tilde{\alpha}=\frac{2\pi R_0 d}{\epsilon_0}\alpha$ and $\tilde{L}=\frac{L}{R_0}$. The natural scale for $\alpha$ arises from thinking about the total boundary energy for a cell of perimeter $2\pi R_0$, which is $2\pi R_0\Gamma=\frac{2}{3}\epsilon_0\tilde{\alpha}$, with $\Gamma=\frac{2}{3}d\alpha$ (Eq.\ \eqref{lin}). Our threshold in Eq.\ \eqref{bet} quantitatively captures the transition between dominances of training and walk-past outcomes, and explains why small $\alpha$, large $\beta$ and large $L$ could more predominantly cause a walk-past. 

Within our linear stability theory, what is the role of cell-cell adhesion?  Until now, we have assumed that the energy of the cell's boundary is solely controlled by the line tension $\Gamma$, without addressing the quantitative effects of the weak cell-cell adhesion in our model ($\sigma$). However, our cell-cell adhesion term in Eq.\ \eqref{fij} will lower the energy of the cell-cell boundary, effectively changing the cell-cell interfacial tension to $\Gamma - c \sigma$, with $c$ some constant (see, e.g.\ \cite{Camley2014}.) We compute this value of $c$ ($c \sigma$ turns out to be small relative to $\Gamma$) and show how this alters our predicted $\beta_\textrm{crit}$ in  Appendix \ref{app:shift}. The effect of the absence of cell-cell adhesion is relatively minor, shifting the transition line leftward to smaller $\alpha$ by an amount of $0.0134\epsilon_0/l_0^2$ in the phase diagrams from Fig.\ \ref{L90} to \ref{L90s0} as a result of changing $\sigma=5\epsilon_0/l_0^2$ (in Eq.\ \eqref{fij}) to $\sigma=0$. We note that this calculation of $c$ is fairly rough, so we see some disagreements with the predicted line in, e.g.\ Fig.\ \ref{L70s0} and \ref{L80s0} with $\sigma=0$ and smaller fiber spacing $L$ where some points dominated by walk-past are below the shifted line. However, our model correctly captures the direction and overall scale of the shift in the phase diagram due to cell-cell adhesion. 

There are some subtleties in our linear stability analysis that we should mention. First, the symmetry-breaking we are describing here must take place in competition with other processes we have not described -- e.g., a repolarization away from the cell-cell contact which could lead to a reversal. Our theory will thus never capture the transition to reversals in Fig.\ \ref{L90s0}. Secondly, we note that the same basic competition between positive feedback driving asymmetry and line tension suppressing it can also be present in a single cell, and not just a cell-cell collision, though there will be less positive feedback, since only one cell will be pushing at the interface. We think this explains the cases where cells are unstable on both fibers (blue dots in Fig.\ \ref{L90}), which will also be more prominent at low $\alpha$ and high $\beta$. We also note that our specific fit parameters, in Eq.\ \eqref{bet}, are tuned to the transition between walk-past and training being predominant, not the first appearance of walk-past. We believe this point is easier to identify in a finite, noisy data set. 

\section{\label{sec:disc} Discussion}

Our simulations and linear stability analysis show that both mechanical factors like the cell tension and biochemical factors like the positive feedback strength, along with the fiber spacing, can control the outcome of cell-cell collisions in the parallel geometry, switching cells between walk-past and training. Experimentally, NIH 3T3 cells in the parallel geometry almost always train, unless they have recently divided, in which case they most commonly walk past one another \cite{Singh2021}. Cells that have recently divided have higher speeds \cite{Singh2021} -- and in our model, cells with higher positive feedback values $\beta$ also have higher cell speeds (see Fig.\ \ref{speed} for a plot of cell speed changing with $\beta$). Our model is in reasonable agreement with this central finding of \cite{Singh2021}. This idea may also be more strongly experimentally tested by looking at, e.g.\ cells with up-regulation of Rac1 activity as a reasonable guess for the relevant positive feedback. Our model could be further challenged by modification of cell tensions \cite{Maitre2012,Lin2023,Yamada2007} to change the deformation energy of the boundary.  We also predict a relatively large effect of the fiber spacing -- a small decrease of inter-fiber separation $L$ (about 22\%) can drastically alter the distribution of dominant outcomes in phase diagrams from Fig.\ \ref{L90} to \ref{L70}. This is possible to test, though not straightforward, as depending on the stiffness of the fibers, pairs of parallel fibers may be pulled together \cite{Singh2021,Sheets2013}. 

One of the most striking common findings in both Singh et al.\ \cite{Singh2021} and our numerical studies is the near absence of the classical CIL -- the mutual reversal, either on one fiber or two fibers. Within our model, reversals are largely present when the cell-cell adhesion is zero, but the cell's positive feedback strength $\beta$ is low and line tension is high -- suppressing walk-past. However, this is also the largest contrast between our results and the experimental results in Singh et al.\ \cite{Singh2021} -- we predict relatively minimal effect of cell-cell adhesion on the transition between training and walk-past (in fact, the absence of adhesion  promotes slightly more training). By contrast, parallel NIH 3T3s exhibit training with normal cell-cell adhesion but would switch to walk-past in calcium-free media with low adhesion. This inconsistency suggests either 1) our model is oversimplified, 2) disruption of E-cadherin also disrupts cell-cell signaling that is necessary for contact interactions \cite{Roycroft2016}, and this also promotes walk-past, or 3) the low-calcium media perturbation used in \cite{Singh2021} also changed another key variable, e.g.\ changing the cell's tension or positive feedback. Calcium naturally plays a large role in many factors of cell migration \cite{Sengupta2021,Hung2016}, and the low-calcium media can therefore affect more than just cell-cell adhesions. 

Our model provides a key new idea -- that walk-past may be controlled by linear stability of the symmetric cell-cell interaction, and thus managed by cell mechanics and environment geometry. These results could be brought into closer connection with experimental work in a few different directions, either via expanding our current \lq\lq bottom-up\rq\rq\ model, or via a data-driven  \lq\lq top-down\rq\rq\ approach \cite{Bruckner2021,Bruckner2024}. It would be interesting to see if our linear stability hypothesis also holds in more realistic models of cell polarity \cite{Camley2014,Kulawiak2016,PerezIpina2024}. However, extension of these models to the fiber geometry may not be straightforward. Our model reproduces spindle and parallel cell shapes similar to experimental observations, but real cell bodies are much more stretched and elongated along fibers  \cite{Sheets2013,Jana2019}. More detailed models of protrusion formation and stabilization, e.g.\ following \cite{Caballero2014,LoVecchio2020}, might allow for improvement in this direction. Another potential path to closer contact with experiment is to connect to data-driven models for cell collision learned from experimental measurements \cite{Bruckner2021,Bruckner2024}.  The measurements of \cite{Bruckner2021} show that noncancerous cells are characterized by more cell-cell repulsion and cell-substrate friction, while metastatic cells feature more cell-cell attraction and anti-friction, leading to more sliding events (walk-past). Connecting these quantitative details to the underlying simple linear stability idea could be a fruitful area for future work. %

\begin{acknowledgments}
The authors acknowledge support from the National Science Foundation award MCB 2119948 (YL and BAC) and MCB 2119949 (ASN). This work was carried out at the Advanced Research Computing at Hopkins (ARCH) core facility (rockfish.jhu.edu), which is supported by the National Science Foundation (NSF) grant number OAC 1920103. We thank Pedrom Zadeh for useful conversations, including feedback on the polarity model we use, originally developed for \cite{Zadeh2024}, and thank Emiliano Perez Ipi\~na and Wei Wang for a careful reading of the draft. 
\end{acknowledgments}

\appendix
\renewcommand\thefigure{\thesection\arabic{figure}}  

\section{\label{app:fig} Supplementary figures}

Additional figures mentioned in the main text, including simulation outcomes of cell collisions at intermediate fiber spacing $L = 80l_0$ (Fig.\ \ref{L80}) and at $L = 70l_0$ and $L = 80l_0$ without cell-cell adhesion (Fig.\ \ref{L70s0} and \ref{L80s0} respectively), and the dependence of model cell speed on positive feedback strength $\beta$ (Fig.\ \ref{speed}), are shown in this section.

\setcounter{figure}{0} 
\begin{figure}[hbt!]
\centering
\includegraphics[width=\columnwidth]{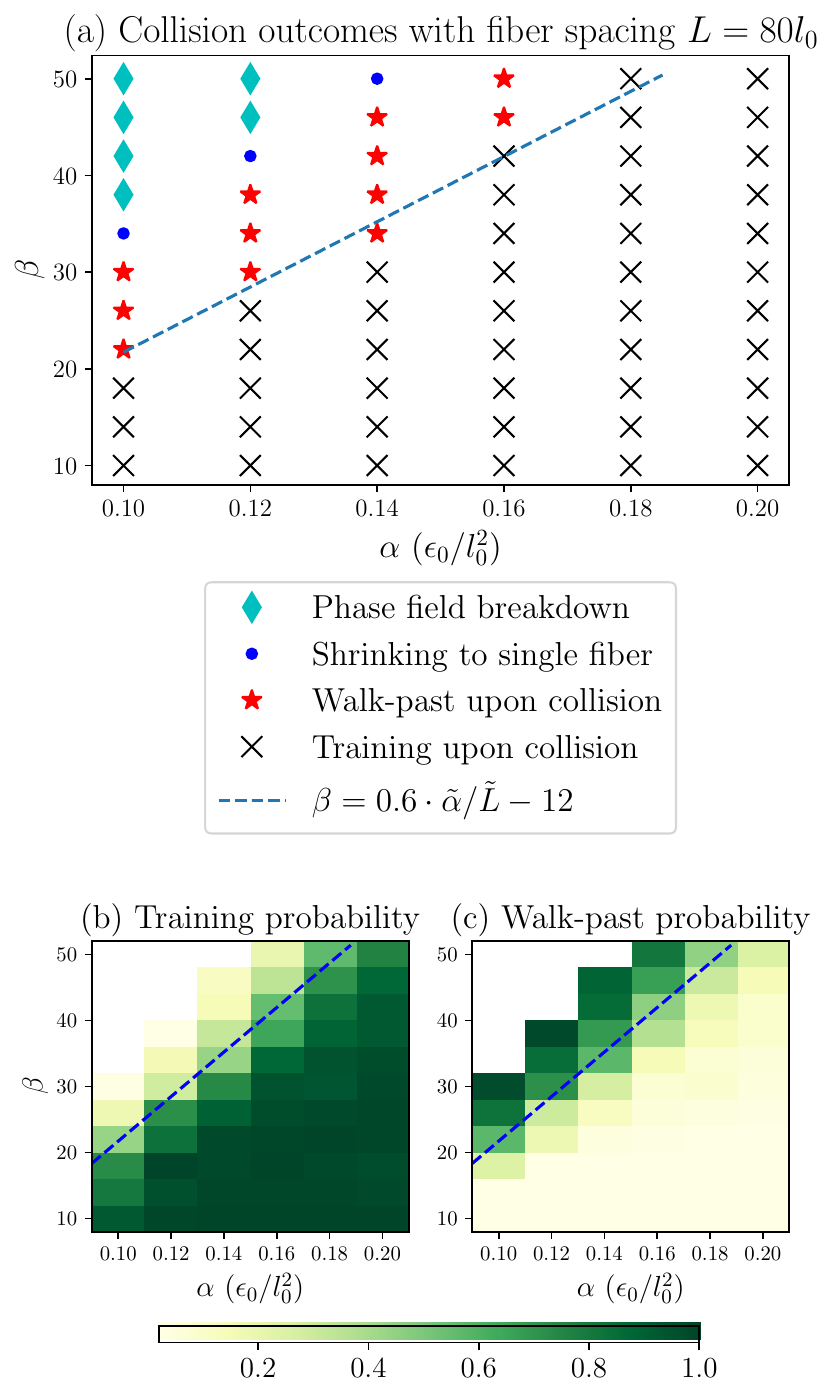}
\caption{(a) Phase diagram showing dominant cell collision outcomes on two parallel fibers separated by a spacing $L=80l_0$ {(\SI{20}{\micro\m})}. (b) and (c) show the frequencies of training and walk-past results calculated from 96 independent simulation trajectories at each point in the phase diagram.}
\label{L80}
\end{figure}

\begin{figure}[hbt!]
\centering
\includegraphics[width=\columnwidth]{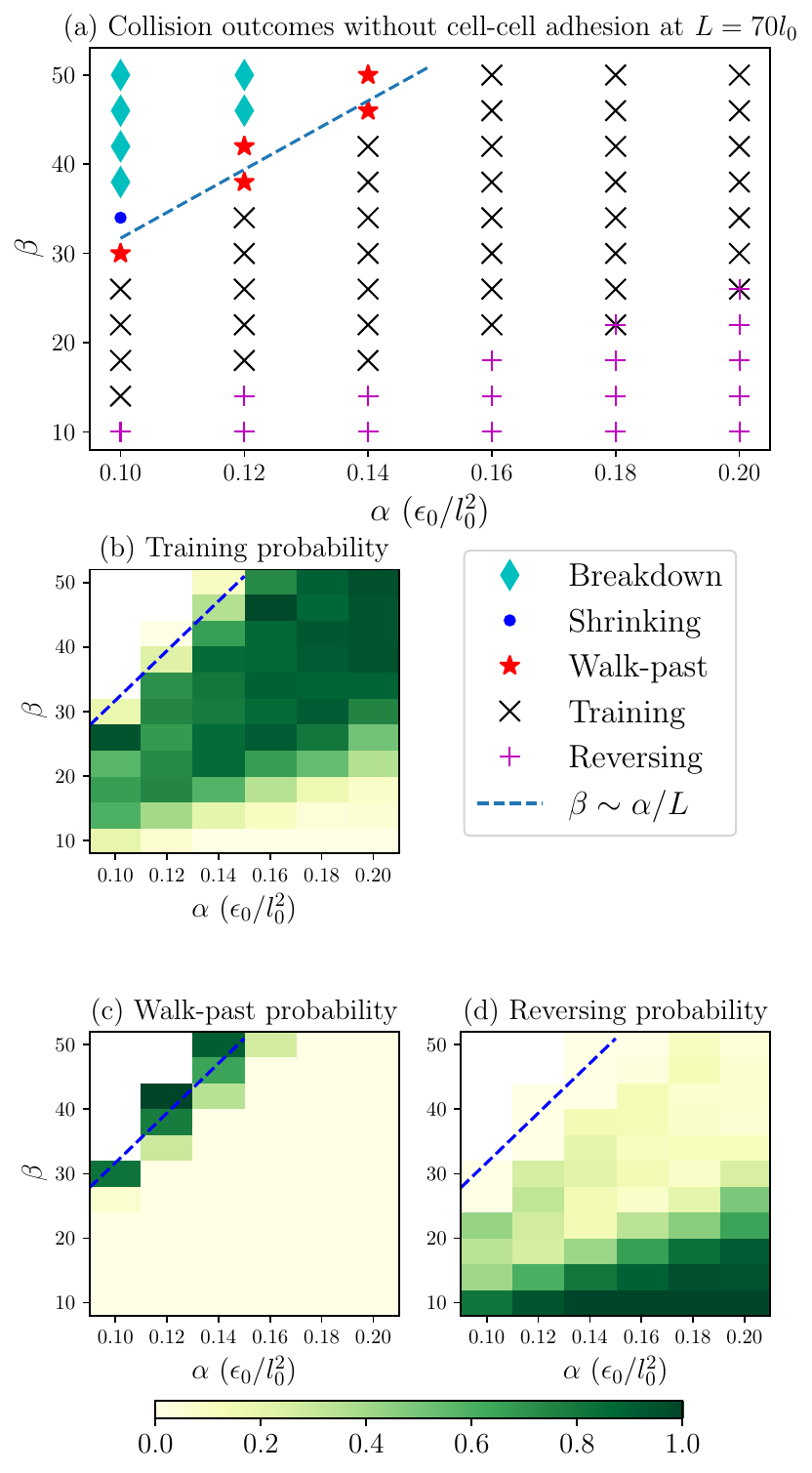}
\caption{(a) Phase diagram showing dominant cell collision outcomes without cell-cell adhesion on two parallel fibers separated by a spacing $L=70l_0$ {(\SI{17.5}{\micro\m})}. (b), (c) and (d) show the frequencies of training, walk-past, and reversing results calculated from 96 independent simulation trajectories at each point in the phase diagram.}
\label{L70s0}
\end{figure}

\begin{figure}[hbt!]
\centering
\includegraphics[width=\columnwidth]{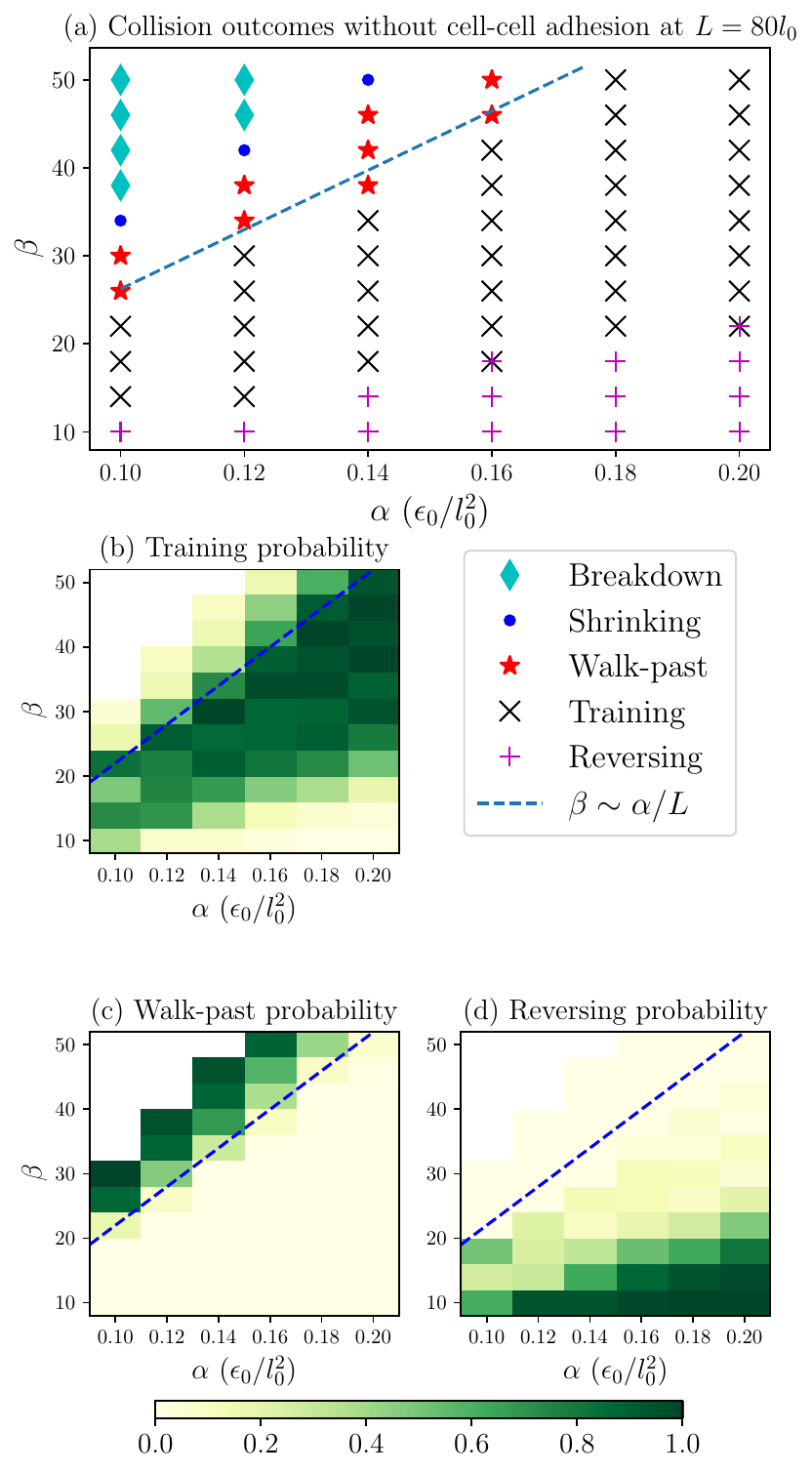}
\caption{(a) Phase diagram showing dominant cell collision outcomes without cell-cell adhesion on two parallel fibers separated by a spacing $L=80l_0$ {(\SI{20}{\micro\m})}. (b), (c) and (d) show the frequencies of training, walk-past, and reversing results calculated from 96 independent simulation trajectories at each point in the phase diagram.}
\label{L80s0}
\end{figure}

\begin{figure}[hbt!]
\centering
\includegraphics[width=\columnwidth]{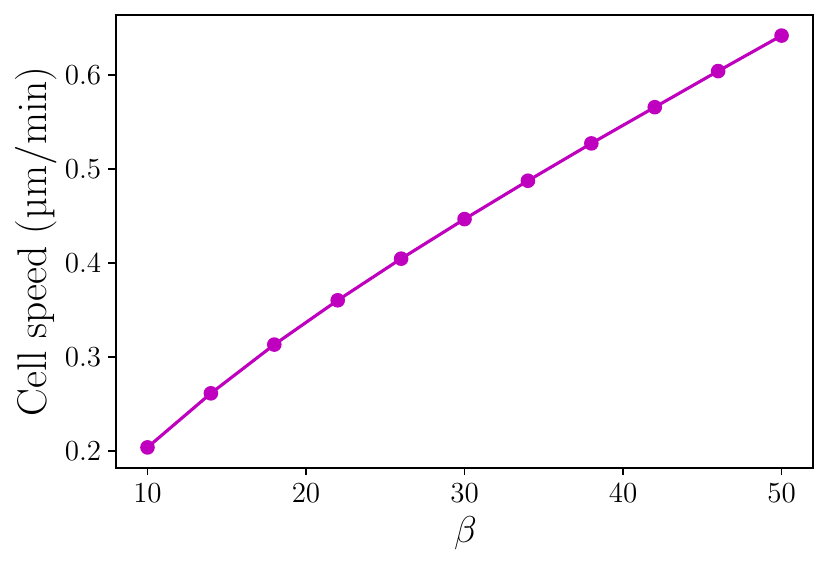}
\caption{Example of cell centroid speed dependent on the positive feedback strength $\beta$ of cell polarity when $\alpha=0.2\epsilon_0/l_0^2$ and fiber spacing $L=90l_0$. Each data point is an average speed from 20 simulations and the standard error is smaller than the marker size.}
\label{speed}
\end{figure}

\section{\label{app:simul} Simulation details}

In our numerical modeling, the 2D simulation space used for the cell phase field $\phi_i$ and its associated polarization field $\mathbb{P}_i$ is discretized with lattice size $l_0=\SI{0.25}{\micro\m}$, and the simulation time step is $\Delta t=0.002\tau_0=\SI{0.001}{\min}$ (see Table \ref{param}). We use a periodic boundary condition for the simulation space, which has a dimension of $200l_0\times200l_0$ for a single cell (Fig.\ \ref{single}) or $400l_0\times200l_0$ for two cells (Fig.\ \ref{double}). At the beginning of each simulation trajectory, $\phi_i$ is initially in a circular shape of radius $R_0=40l_0$, which is then placed on top of the fiber substrate field $\chi(\vek{r})$ (consisting of either a single fiber or double parallel fibers whose width is $2l_0$) and relaxed without active forces. We use a forward in time, central in space (FTCS) scheme to calculate the spatiotemporal evolution of phase field $\phi_i$:
\begin{equation}
\label{phi}
\phi_i(\vek{r}, t+\Delta t)=\phi_i(\vek{r}, t)+\frac{\partial\phi_i}{\partial t}(\vek{r}, t)\Delta t
\end{equation}
where the time derivative $\partial\phi_i/\partial t$ is calculated at $\vek{r}=(x,y)$ (position on the numerical lattice) and time $t$ (using Eq.\ \eqref{evo}) from the instant evaluation of:
\begin{equation}
\label{dfdp}
\begin{split}
\frac{\delta F_i}{\delta\phi_i}=
&\alpha(\phi_i^3-\frac{3}{2}\phi_i^2+\frac{1}{2}\phi_i-\frac{d^2}{2}\nabla^2\phi_i)\\
&-\frac{4\lambda\phi_i}{\pi R_0^2}\Big(1-\int\dif\vek{r}\frac{\phi_i^2}{\pi R_0^2}\Big)-\frac{6E_0 A_{\text{ref}}\phi_i(1-\phi_i)\chi}{(A_{\text{ref}}+A_{\text{ad}}(\phi_i))^2}
\end{split}
\end{equation}
for a single cell $i$ without interacting with other cells, and
\begin{equation}
\begin{split}
\nabla^2\phi_i(x,y)=
&\frac{1}{l_0^2}\Big[\phi_i(x+l_0,y)+\phi_i(x-l_0,y)\\
&+\phi_i(x,y+l_0)+\phi_i(x,y-l_0)-4\phi_i(x,y)\Big]
\end{split}
\end{equation}
\begin{equation}
\label{grad}
\begin{split}
\nabla\phi_i(x,y)=
&\Big(\frac{\phi_i(x+l_0,y)-\phi_i(x-l_0,y)}{2l_0},\\
&\frac{\phi_i(x,y+l_0)-\phi_i(x,y-l_0)}{2l_0}\Big)
\end{split}
\end{equation}
all at time $t$. The calculation of $\vek{v_i}(\vek{r}, t)$, which is needed for calculating $\partial\phi_i/\partial t$, is done by using Eq.\ \eqref{vi}, which also uses the calculation of Eq.\ \eqref{dfdp} and \eqref{grad} and implements the random noise $\vek{\tilde{v}_i}$ every $t_\textrm{sample} = 0.002\tau_0$ (with the noise generated at each sampling time point and kept being applied for a period of $t_\textrm{sample}$ each time). Here $t_\textrm{sample}$ is chosen to be equal to $\Delta t$, so that the velocity noise is sampled and added at each simulation time step and updated at the next time step in our algorithm. We note that changing the time step $\Delta t$ should not vary the noise application timescale $t_\textrm{sample}$, and would require changing the frequency and duration of the application of the velocity noise. For instance, if $\Delta t$ is changed to $\Delta t_\text{new}=\Delta t/n$ then the noise should be sampled every $t_\textrm{sample} = n\Delta t_\text{new}$, after which point the same noise also persists for $n\Delta t_\text{new}$ until being updated at the next sampling time point. This gives a consistent perturbation onto cell velocity. We mention this only for completeness -- we have presented no results where the time step is changed. 

When we initialize our simulations, we first relax the cell's shape by evolving it with no active motility force, i.e.\ $\vek{f_i}=0$.  This cell shape relaxation lasts for a duration of $40\tau_0$, resulting in a static spindle shape (on one fiber) or a parallel shape (on two fibers) as in the left column of Fig.\ \ref{single}. 

After the initial static relaxation period, an asymmetric polarization field is applied to the cell, kick-starting the positive feedback and providing a driving force for cell motion. For example, for a cell to start moving rightward, we set $\mathbb{P}_i=1$ in the right half of the initially static, symmetric cell, and $\mathbb{P}_i=0$ in the left half of the cell (and also outside the cell). The polarization field $\mathbb{P}_i$ evolves with time through:
\begin{equation}
\label{pi}
\mathbb{P}_i(\vek{r}, t+\Delta t)=\Big(\mathbb{P}_i(\vek{r}, t)+\frac{\partial\mathbb{P}_i}{\partial t}(\vek{r}, t)\Delta t\Big)\Theta(\phi_i(\vek{r}, t)-\phi_{\text{min}})
\end{equation}
in which $\Theta(\phi_i-\phi_{\text{min}})$ is a Heaviside step function equal to one wherever $\phi_i>\phi_{\text{min}}=0.001$ and zero where $\phi_i<\phi_{\text{min}}=0.001$. This sets $\mathbb{P}_i = 0$ far from the cell. The time derivative $\partial\mathbb{P}_i/\partial t$ is evaluated at $\vek{r}$ and $t$ by using Eq.\ \eqref{ppi}, which requires the calculation of $\partial\phi_i/\partial t$ using Eq.\ \eqref{evo} (which involves Eq.\ \eqref{dfdp}--\eqref{grad} as we have illustrated above). Here the nonzero cell polarity gives rise to a motility force field $\vek{f_i}(\vek{r}, t)$ calculated by Eq.\ \eqref{ffi} using the current $\phi_i(\vek{r}, t)$ and $\mathbb{P}_i(\vek{r}, t)$. $\vek{f_i}(\vek{r}, t)$ is further used in Eq.\ \eqref{vi} to obtain $\vek{v_i}(\vek{r}, t)$, which is substituted into Eq.\ \eqref{evo} for the evaluation of $\partial\phi_i/\partial t$ which is then used in both Eq.\ \eqref{phi} and \eqref{ppi} (and then \eqref{pi}) to obtain  $\phi_i(\vek{r}, t+\Delta t)$ and $\mathbb{P}_i(\vek{r}, t+\Delta t)$.

As the dynamic simulation proceeds, the freely migrating cell takes an asymmetric shape (as in the middle column of Fig.\ \ref{single}) and the polarization field takes the value of $\mathbb{P}_i>0$ near the cell front and $\mathbb{P}_i<0$ near the rear (as in the right column of Fig.\ \ref{single}). Both cell shape and polarization field become stabilized in steady cell motion before the end of this free motion period of $400\tau_0$ (unless the phase field breaks down or parallel cell becomes unstable and shrinks to single fiber as in the upper left corner of each phase diagram in Fig.\ \ref{L90}--\ref{L90s0} (a), for which collision is not simulated). In the last simulation period afterwards (cell collision), we place two identical cells moving in opposite directions (already in steady motion, with phase and polarization fields generated in identical conditions from the previous dynamic stabilization period) together in the same simulation space (as in the left column of Fig.\ \ref{double}) and make them continue moving and experience a head-to-head collision. The presence of cell-cell interaction requires using the functional derivative of the full free energy $\delta F/\delta\phi_i$ (not just $\delta F_i/\delta\phi_i$ for single cell in Eq.\ \eqref{dfdp}):
\begin{equation}
\frac{\delta F}{\delta\phi_i}=\frac{\delta F_i}{\delta\phi_i}+2g\phi_i\phi_j^2-2\sigma(\phi_i^3-\frac{3}{2}\phi_i^2+\frac{1}{2}\phi_i)\phi_j^2(1-\phi_j)^2
\end{equation}
for $j\neq i$. We perform the simulation for another $400\tau_0$, and observe the collision outcomes by keeping track of both cells using their centers of mass, which we estimate by using the method introduced in \cite{Bai2008} in a periodic boundary condition.

\section{\label{app:shift} Effect of cell-cell adhesion in the linear stability theory}
\setcounter{figure}{0} 
\begin{figure}[hbt!]
\centering
\includegraphics[width=\columnwidth]{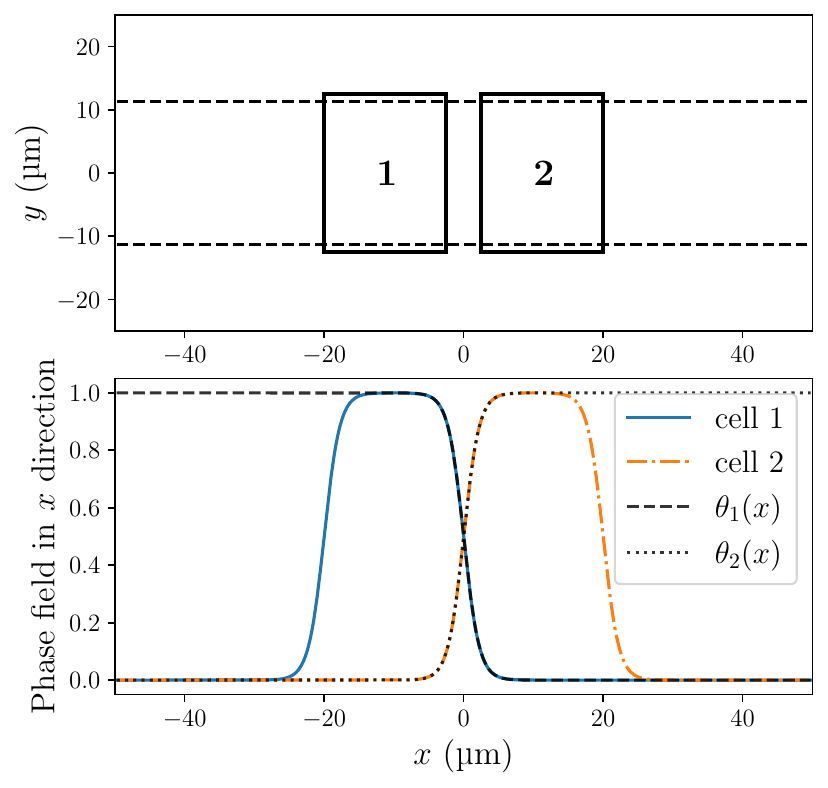}
\caption{Example of cell-cell contact in 2D (upper panel) and 1D (lower panel).}
\label{adh}
\end{figure}

To estimate the quantitative effect of the adhesion between cell boundaries in touch, specifically how much it affects the location of the transition line in the phase diagrams between dominantly walk-past and training as collision outcomes predicted from the linear stability analysis, we start by considering the simplified cell shapes made of straight-line boundary segments in the upper panel of Fig.\ \ref{adh} (similar to those in Fig.\ \ref{lst} (b)) of two cells coming into contact with one another at horizontal position $x=0$. When crossing a cell boundary in our simulations, the 2D phase field of a model cell transitions between $\phi=1$ and $\phi=0$ in a smooth way (across a boundary width $d=\sqrt{20}l_0$) that can be described by a hyperbolic $\tanh$ function. In the example of Fig.\ \ref{adh}, which shows the 1D phase field profiles of cells 1 and 2 along the $x$ direction (as cross sections of their 2D phase fields) in the lower panel, functions $\theta_1(x)=\frac{1}{2}+\frac{1}{2}\tanh(-\frac{x}{2d})$ and $\theta_2(x)=\frac{1}{2}+\frac{1}{2}\tanh(\frac{x}{2d})$ are used to represent the right boundary of cell 1 and the left boundary of cell 2, respectively, which are in contact at $x=0$ (where $\theta_1(0)=\theta_2(0)=0.5$). We suppose the interface formed between the cells has a length of $l_{\text{int}}$, and using the second half of Eq.\ \eqref{fij}, we calculate the contribution of cell-cell adhesion to the total free energy (lowering of free energy) for each cell to be:
\begin{equation}
\begin{split}
\Delta E_{\text{adh}}
&=-\frac{\sigma}{4}\int\dif\vek{r}\phi_1^2(1-\phi_1)^2\phi_2^2(1-\phi_2)^2\\
&\approx -\frac{\sigma}{4}l_{\text{int}}\int\dif x\theta_1^2(1-\theta_1)^2\theta_2^2(1-\theta_2)^2\\
&=-0.008\sigma l_0 l_{\text{int}}.
\end{split}
\end{equation}
This means that -- from the standpoint of our linear stability theory -- that the cell-cell adhesion will act as a negative line tension. Though we derived this for two cells with a straight cell-cell boundary, we expect this expression can be used to estimate the adhesion energy between phase field cells with an interface of length $l_{\text{int}}$ in any shape as long as the interface thickness $d$ times the curvature of the boundary is sufficiently small \cite{elder2001sharp}. For the same interface, the energetic contribution from line tension $\Gamma$ is $\Gamma l_{\text{int}}=\frac{2}{3}d\alpha l_{\text{int}}$ according to Eq.\ \eqref{lin}. 

To quantify the amount of shift of the location of linear stability prediction (dashed threshold) in the phase diagrams, we combine both line tension and adhesion effects and find that by having $\Delta E_{\text{adh}}=\Delta\Gamma_{\text{adh}} l_{\text{int}}$ (or $-0.008\sigma l_0 l_{\text{int}}=\frac{2}{3}d\Delta\alpha_{\text{adh}} l_{\text{int}}$), the cell-cell adhesion effectively changes the line tension to $\Gamma+\Delta\Gamma_{\text{adh}}=\Gamma-c\sigma$ where $c=0.008l_0$, and contributes an amount of $\Delta\alpha_{\text{adh}}=-0.0134\epsilon_0/l_0^2$ to the parameter $\alpha$ given other parameters in Table \ref{param}. 
When the adhesion is turned off by changing $\sigma=5\epsilon_0/l_0^2$ to $\sigma=0$, the threshold moves leftward horizontally to smaller $\alpha$ by $0.0134\epsilon_0/l_0^2$.

\section{Movie captions}

Movie 1: A typical collision between spindle-shaped cells on a single fiber, which almost always results in training. Here we use parameters $\alpha=0.1\epsilon_0/l_0^2$ and $\beta=10$.%

Movie 2: A collision between parallel-cuboidal cells on two parallel fibers, which results in a training. Here we use parameters $\alpha=0.1\epsilon_0/l_0^2$ and $\beta=10$, and fiber spacing $L=\SI{22.5}{\micro\m}$.

Movie 3: An example of walk-past between parallel cells upon collision on two fibers. Here we use parameters $\alpha=0.1\epsilon_0/l_0^2$ and $\beta=14$, and fiber spacing $L=\SI{22.5}{\micro\m}$ (here walk-past is a possible but not dominant outcome). Both cells shrink to different single fibers (with each cell randomly shrinking to one of the two fibers) and become teardrop-shaped during the walk-past.

Movie 4: An example of two-cell reversing (like a classical CIL) upon collision on two parallel fibers. Here we use parameters $\alpha=0.1\epsilon_0/l_0^2$ and $\beta=10$, and fiber spacing $L=\SI{22.5}{\micro\m}$ (here reversal is a possible but not dominant outcome). Both cells keep their parallel shapes after collision.

Movie 5: An example of collision between parallel cells on two fibers, which results in a training during which one cell shrinks to a single fiber. Here we use parameters $\alpha=0.12\epsilon_0/l_0^2$ and $\beta=14$, and fiber spacing $L=\SI{22.5}{\micro\m}$.

Movie 6: An example of collision between parallel cells on two fibers, which results in a reversal during which one cell shrinks to a single fiber. Here we use parameters $\alpha=0.14\epsilon_0/l_0^2$ and $\beta=14$, and fiber spacing $L=\SI{22.5}{\micro\m}$ (here reversal is a possible but not dominant outcome).

Movie 7: An example of collision between parallel cells on two fibers, which results in a training during which both cells shrink to single fibers. Here we use parameters $\alpha=0.1\epsilon_0/l_0^2$ and $\beta=14$, and fiber spacing $L=\SI{22.5}{\micro\m}$.

Movie 8: An example of collision between parallel cells on two fibers, which results in a reversal during which both cells shrink to single fibers. Here we use parameters $\alpha=0.1\epsilon_0/l_0^2$ and $\beta=14$ \emph{as well as zero cell-cell adhesion}, and fiber spacing $L=\SI{22.5}{\micro\m}$ (here reversal is a possible but not dominant outcome).

Movie 9: A single cell initially attached to two parallel fibers becomes unstable and shrinks to a single fiber on its own under strong driving force. Here we use parameters $\alpha=0.1\epsilon_0/l_0^2$ and $\beta=30$, and fiber spacing $L=\SI{22.5}{\micro\m}$.

\bibliography{YL_cell_collision}%

\end{document}